\documentclass[aps,prd,twocolumn,showpacs,preprintnumbers,superscriptaddress,nofootinbib]{revtex4-1}
\usepackage{amssymb}
\usepackage{bm}
\usepackage{enumerate}
\usepackage{amssymb}
\usepackage{amsmath}
\usepackage{feynmf}
\usepackage{slashed}
\usepackage{wrapfig}
\usepackage{graphicx}

\newcommand\be{\begin{equation}}
\newcommand\ee{\end{equation}}

\newcommand\bea{\begin{eqnarray}}
\newcommand\eea{\end{eqnarray}}

\begin{document}

\title{Cosmological constraints from CMB distortion}

\author{James B.\ Dent}
\email{jbdent@asu.edu}
\affiliation{Department of Physics \& School of Earth and Space Exploration, Arizona State University, Tempe, AZ 85287-1404}

\author{Damien A. Easson}
\email{easson@asu.edu}
\affiliation{Department of Physics \& School of Earth and Space Exploration, Arizona State University, Tempe, AZ 85287-1404}

\author{Hiroyuki Tashiro}
\email{hiroyuki.tashiro@asu.edu}
\affiliation{Department of Physics  \& School of Earth and Space Exploration, Arizona State University, Tempe, AZ 85287-1404}

\begin{abstract}
We examine bounds on adiabatic and isocurvature density fluctuations from $\mu$-type spectral distortions of the cosmic microwave background (CMB).
Studies of such distortion are  complementary to CMB measurements of the spectral index
and its running, and will help to constrain these parameters on significantly smaller scales.   We show that a detection on the order of $\mu \sim 10^{-7}$ would
strongly be at odds with the standard cosmological model of a nearly scale-invariant spectrum of adiabatic perturbations. Further, we find that
given the current CMB constraints on the isocurvature mode amplitude,  a nearly scale-invariant isocurvature mode (common in many curvaton models)
cannot produce significant $\mu$-distortion. Finally, we show that future experiments will strongly constrain the amplitude of the isocurvature modes with a highly blue spectrum as predicted by certain axion models. 

\end{abstract}



\maketitle 

\section{Introduction}

The scale-invariant formulation of the power spectrum of 
primordial fluctuations is highly successful
\cite{Komatsu:2010fb}. The observational results of both cosmic microwave background (CMB) and large
scale structure (LSS) studies are modeled superbly by a scale-invariant
spectrum.  However,  observational data for the power spectrum on
small scales (smaller than 0.1 Mpc) is lacking, prompting
discussions of a possible deviation from scale invariance on
small scales. For example, many inflationary models predict a running
spectral index, which may develop a large deviation from a scale
invariant spectrum  at large CMB multipoles $\ell$  (for a survey of theoretical
motivations and observational prospects for a running index, see
\cite{Adshead:2010mc}).

Isocurvature fluctuation modes can also produce
deviations from a scale invariant spectrum.  For example, isocurvature
fluctuations are generated in the axion dark matter model \cite{axion}
and the curvaton scenario \cite{curvaton}.  If the spectrum is blue,
the isocurvature modes contribute additional power to the density
fluctuations on small scales, although the isocurvature mode remains
subdominant on large scales.  Current observational constraints
allow models with a mixture of scale-invariant adiabatic and
isocurvature modes with a blue spectrum \cite{isoconstraint}, with
isocurvature modes constrained to contribute at most 10\% on large scales
(pure isocurvature models are already observationally ruled out \cite{enqvist:2002}).

Currently, the running of the power spectrum index for adiabatic modes and
the amplitude of isocurvature modes are constrained by CMB and large
scale structure \cite{Komatsu:2010fb}. For the running power spectrum,
the best fit value for WMAP and large scale structure is $-0.022 \pm
0.020$. Furthermore, the amplitude ratio of the
isocurvature mode to the adiabatic mode at $k=0.002 {\rm Mpc}^{-1}$ is
0.13 and 0.011 for the uncorrelated CDM isocurvature case and the
correlated one, respectively.  However these constraints can be vastly
improved by measurements of the fluctuations on smaller scales.

The measurement of the CMB spectrum distortion from the perfect blackbody
shape is a powerful tool for producing constraints on small scale
density fluctuations before the recombination epoch. The energy of the
density fluctuation on small scales is released into CMB photons via Silk damping \cite{Silk:1967kq}. 
At a very high redshift
($z>10^6$), the released energy is quickly thermalized by
double-Compton and Compton scattering, and the spectrum of the CMB
photons maintains a blackbody shape. However, as the universe
evolves, these scatterings become less effective and cannot establish
 thermalization of the released energy. As a result, released
energy is imprinted on the CMB spectrum as  CMB distortions from
the blackbody shape \cite{Daly:1991,hu:1994}. Such distortions may be classified
into two types, $\mu$-distortion and $y$-distortion
\cite{hu:1993}. While $\mu$-distortion is produced from the double
Compton scattering decoupling ($z \sim 10^6$) to the thermalization
decoupling by Compton scattering ($z \sim 10^5$), $y$-distortion is
generated from thermalization decoupling by Compton scattering to the
recombination epoch. In other words, by measuring these distortions, we
can learn about the density fluctuations on scales smaller than the Silk
damping scale at these redshifts, up to $k \sim 10^{5} {\rm Mpc}^{-1}$.

The current constraints on these distortions, obtained by
COBE FIRAS, are $|\mu|<9 \times 10^{-5}$ and $y< 1.5 \times 10^{-5}$
\cite{cobefiras}.  Proposals for future space missions such as PIXIE
have the potential to provide dramatically tighter constraints on both
types of distortion, with projected detection levels of $|\mu| \sim
5 \times 10^{-8}$ or $y \sim 10^{-8}$ by PIXIE at the 5 $\sigma$ level
\cite{pixie}.  It was recently pointed out that PIXIE can detect the
CMB distortion due to the dissipation of the primordial fluctuations
with $n_s= 0.96$ at a 1 $\sigma$ level, because the distortions for
$n_s= 0.96$ are $\mu \sim 8 \times 10^{-9}$ \cite{Chluba:2011hw,Khatri:2011aj,Chluba:2012gq}.

In this paper, we investigate $\mu$-distortions caused by the two types of
fluctuations described above. One is the primordial fluctuation with
 running spectral index and the other is an additional power-law
fluctuation which is sub-dominant on large scales.  One of the
motivations for the latter type is the CDM or baryon isocurvature mode
with a blue spectrum.  An extremely blue spectrum ($n_{iso} >2$) can
be generated in the axion model suggested in \cite{kasuya:2009}.
Although the density fluctuations of CDM and baryons due to the
isocurvature mode can survive Silk damping, the photon density
fluctuations produced by the isocurvature mode suffer Silk damping
\cite{hu-sugiyama:1995} and this dissipation can produce
$\mu$-distortion.

\section{Distortions from Curvature Perturbations}

The energy injection in the early universe produces  CMB spectral
distortions. The evolution of the spectral distortions depends
on the energy injection rate and the time  scale of the thermalization
process. In particular, in the regime between $z \sim 10^6$ and
$z \sim 10^5$, $\mu$-distortion is created by the balance
between the energy injection and the double Compton scattering.
The time evolution of $\mu$-distortion due to this energy injection is
given by \cite{hu:1994}
\begin{eqnarray}
\frac{d\mu}{dt} = -\frac{\mu}{t_{DC}(z)} + 1.4\frac{dQ/dt}{\rho_{\gamma}},
\end{eqnarray}
where $t_{DC}$ is the time scale for double Compton scattering
\begin{eqnarray}
t_{DC} = 2.06\times 10^{33} \left(1-\frac{Y_p}{2} \right)^{-1}(\Omega_bh^2)^{-1}z^{-9/2}s,
\end{eqnarray}
where $Y_p$ is the primordial helium mass fraction.  The solution to this evolution is
\begin{eqnarray}\label{mu}
\mu &=& 1.4\int_{t(z_1)}^{t(z_2)}dt\frac{dQ/dt}{\rho_{\gamma}}e^{-(z/z_{DC})^{5/2}} \nonumber \\
&=& 1.4\int_{z_1}^{z_2}dz\frac{dQ/dz}{\rho_{\gamma}}e^{-(z/z_{DC})^{5/2}}, 
\end{eqnarray}
where
\begin{eqnarray}
z_{DC} = 1.97 \! \times \! 10^6\left(1-\frac{1}{2}\left(\frac{Y_p}{.24}\right)\right)^{-2/5} \! \left(\frac{\Omega_bh^2}{.0224}\right)^{-2/5}
\end{eqnarray}

The dissipation of acoustic waves in the photon-baryon plasma due to
Silk damping injects  energy into the CMB and causes the CMB spectral
distortions.

We can write the energy density perturbation of the acoustic waves as\footnote{This
form is for the case of non-relativistic particles, whereas the full expression may be found in
\cite{Khatri:2011aj,Khatri:2012rt}.  We thank R. Khatri for pointing this out to us.}
\begin{equation}
Q \simeq { \rho_\gamma \over 2} \left[
 c_s^2 \langle \delta_\gamma({\bf x})^2 \rangle + \langle v ({\bf
 x}) ^2\rangle \right].
\end{equation}
The density and velocity perturbation of photons are
\begin{eqnarray}
\langle\delta_{\gamma}({\bf x})^2\rangle &=& \int \frac{d^3k}{(2\pi)^3}P_{\gamma}(k)\\
\langle v_\gamma ({\bf x})^2\rangle &=& \int \frac{d^3k}{(2\pi)^3}P_{v}(k).
\end{eqnarray}
The power spectrum $P_{i}(k)$, where the subscript $i$ represents $\gamma$ and
$v$, is related to the primordial power spectrum $P_{\gamma}^i(k)$ by
\begin{eqnarray}
P_{i}(k) =\Delta_{i}^2(k)P_{\gamma}^i,
\end{eqnarray}
where the transfer function $\Delta_{i}(k)$ for modes well inside the horizon satisfies the relation
\begin{eqnarray}
\Delta_{\gamma}(k)\approx 3\textrm{cos}(kr_s)e^{-k^2/k_D^2},
\end{eqnarray}
and
\begin{eqnarray}
\Delta_{v}(k)\approx 3 c_s  \textrm{sin}(kr_s)e^{-k^2/k_D^2},
\end{eqnarray}
with
the sound horizon $r_s$ given by
\begin{eqnarray}
r_s(z) &=&
  \frac{2}{3}\frac{1}{k_{eq}} \sqrt{\frac{6}{R(z_{eq})}}   \nonumber \\
& \times & \ln\left(\frac{\sqrt{1+R(z)}+\sqrt{R(z_{eq})+R(z)}}{\sqrt{R(z_{eq})}+1}\right),
\end{eqnarray}
and $k_D$ is the diffusion scale
\begin{eqnarray}
\frac{1}{k_D^2} = \int_z^{\infty}dz\frac{c(1+z)}{6H(1+R)n_e\sigma_T}\left(\frac{R^2}{1+R}+\frac{16}{15}\right).
\end{eqnarray}

In the above, $R$ is the baryon energy density
\begin{eqnarray}
R \equiv \frac{3\rho_b}{4\rho_{\gamma}},
\end{eqnarray}
$n_e$ is the free electron number density (before the recombination
epoch, it is given by $n_e(z) = (n_{H0} + 2n_{He0})(1+z)^3\equiv
{n_0}(1+z)^3$), and $\sigma_T$ is the Thomson scattering cross-section. 
During radiation domination the diffusion scale becomes 
\begin{eqnarray}
k_D &=& A_D^{-1/2}(1+z)^{3/2}\,\,\,;\,\,\,  \nonumber \\
A_D &=& \frac{8c}{135 H_0\Omega_r^{1/2}n_{e0}\sigma_T} = 5.92\times 10^{10} \textrm{Mpc}^{2}.
\end{eqnarray}
Now,
we can write the energy density in acoustic waves is
\begin{equation}
Q= \Delta _Q (k)^2 P_{\gamma}^i,
\end{equation}
where
\begin{equation}
 \Delta_Q (k) = {3 c_s \over \sqrt{2}} e^{-k^2/k_D^2}.
\end{equation}

The primordial power spectrum can then be related to the power
spectrum for the curvature in the co-moving gauge, $\zeta$, via \cite{Khatri:2011aj} 
\begin{eqnarray}
P_{\gamma}^i = \frac{4}{(\frac{2}{5}R_{\nu} + \frac{3}{2})^2}P_{\zeta}\approx 1.45P_{\zeta}
\end{eqnarray}
where $R_{\nu}$ is the neutrino energy density 
\begin{eqnarray}
R_{\nu} = \frac{\rho_{\nu}}{\rho_{\gamma} + \rho_{\nu}}\approx .4 \,.
\end{eqnarray}
The power spectrum of the curvature is parameterized as \cite{Komatsu:2010fb}
\begin{eqnarray}
P_{\zeta} = \frac{A_{\zeta}2\pi^2}{k^3}\left(\frac{k}{k_0}\right)^{n_s -1 + \frac{1}{2}\textrm{ln}(\frac{k}{k_0})\frac{dn_s}{d \textrm{ln}k}}
\end{eqnarray}
with  normalized amplitude $A_{\zeta} = 2.4\times 10^{-9}$, and pivot
scale $k_0 = .002 \textrm{Mpc}^{-1}$, the spectral index is $n_S$,
and the running of the spectral index is  $\alpha_s \equiv
dn_s/d\textrm{ln}k$.

The energy release per unit redshift is given by
\begin{eqnarray}\label{dQ/dz}
\frac{dQ/dz}{\rho_{\gamma}}=-\int\frac{d^3k}{(2\pi)^3}P_{\gamma}^i(k)\frac{d\Delta_{Q}^2}{dz}  \,.
\end{eqnarray}

\begin{figure}
\begin{center}
\includegraphics[width=0.45\textwidth]{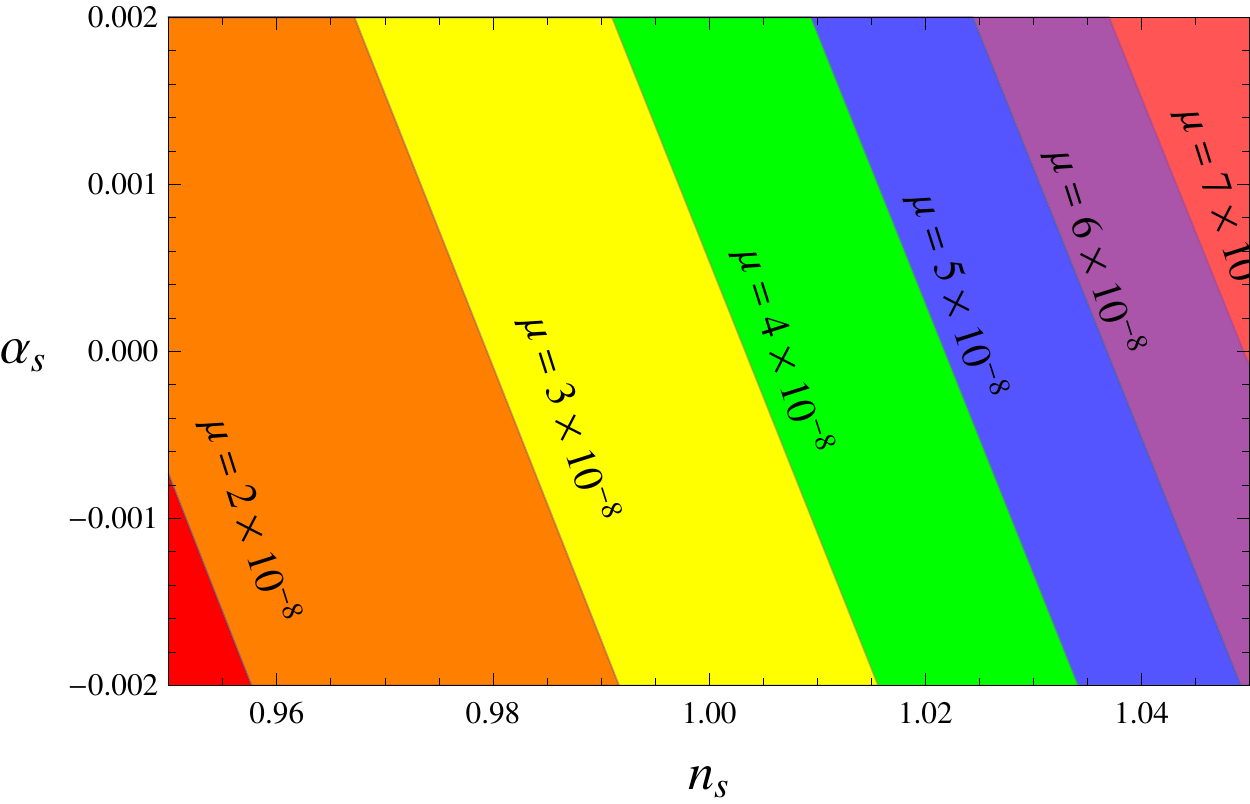}
\end{center}
\caption{Values of $\mu$ generated by energy injection due to the damping of
  acoustic oscillations in the $\alpha_s$-$n_s$ plane for adiabatic modes with amplitude $2.4\times 10^{-9}$.}
\label{fig:mucontours}
\end{figure}

We then solve Eq.(\ref{mu}) while scanning over the $\alpha_s$ - $n_S$
plane using the values $Y_p = .24$, $\Omega_b h^2 = .0224$, along with
the previously mentioned values for $A_{\zeta}$, $A_{D}$, and
$t_{DC}$.  We have calculated $\mu$ for the injection interval $z_1 =
2\times 10^6$ to $z_2 = 5\times 10^4$, and the results are displayed
in Fig.(\ref{fig:mucontours}).

The current WMAP7 bounds are given in Table 7 of
\cite{Komatsu:2010fb}.  Roughly, for no tensor modes and no running ($\alpha_s$ = 0),
$n_s = .968\pm .012$; including running  they find
$n_s = 1.008\pm .042$ with $\alpha_s = -.022\pm.02$.  Including tensor
modes gives $n_s = 1.07\pm .06$, $\alpha_s = -.042\pm.024$, and $r <
.49$ (95\%CL).  The running we have plotted in
Fig.(\ref{fig:mucontours}) is significantly smaller, showing the
constraining ability of such a precise $\mu$ measurement.

With regards to a measurement of the running, $\alpha_s$, Planck may
be able to reach sensitivities on the order of $|\alpha_s| \simeq
.005$, and we see that measurements of $\mu$ at the level of
$10^{-8}$, in tandem with complementary measurements of $n_s$, would
be at least competitive with such a measurement.

As we have stated above, the measurement of $\mu$ may be comparable to or
possibly supersede bounds on the spectral index or its running given
by CMB observations such as WMAP and Planck.  However, it should be
emphasized that the $\mu$ measurement would give information about
such cosmological parameters on much smaller scales, and from effects
occurring at much higher redshifts. Thus, making such measurements is quite
attractive, and would allow us to reach previously unexplored distance and time scales
for cosmological observations.
%


\section{Distortions from Isocurvature Perturbations}

The existence of  additional density fluctuations
in the early universe is another possible heat source
for  the spectral distortion. Candidates of such additional
density fluctuations are isocurvature perturbations.

We define the isocurvature perturbations between photons and
 CDM (or baryons) as
\begin{equation}
 S \equiv {\delta_i } -{3 \over 4} \delta_\gamma,
\end{equation}
where the subscript $i$ is $c$ for the CDM isocurvature and $b$ for
baryon isocurvature.
We parameterized the power spectrum of $S$ similarly to that of
the curvature perturbations
\begin{eqnarray}
P_{iso} = \frac{A_{iso}2\pi^2}{k^3}\left(\frac{k}{k_0}\right)^{m_{iso} -1}
\end{eqnarray}
where we have neglected the possibility of a running isocurvature spectral index.
The value of the amplitude, $A_{iso}$, is constrained by CMB data to
be $A_{iso} \lesssim .1 A_{\zeta}$.

The power spectrum of the isocurvature perturbations for photon density
fluctuations or velocity is given by
\begin{equation}
P_{i}(k) =\Delta_{i,iso}^2(k)P_{iso}(k),
\end{equation}
where $\Delta_{i iso}$ which can be written, well inside the horizon, as
\cite{hu-sugiyama:1995},
\begin{eqnarray}
\label{eq:iso-delta}
\Delta_{\gamma,iso}(k)&\approx& -\sqrt{6} \left(k_{eq} \over k \right)
 \sin (kr_s)e^{-k^2/k_D^2}
\nonumber \\
 \Delta_{v,iso}(k) &\approx& -\sqrt{6} c_s \left(k_{eq} \over k \right)
 \cos (kr_s)e^{-k^2/k_D^2}
\end{eqnarray}
where $k_{eq}$ is the wave number cosseponding to the Hubble horizon
at the equality time.  The energy release per unit redshift is given by
\begin{eqnarray}\label{dQ/dz}
\frac{dQ/dz}{\rho_{\gamma}}=-\int\frac{d^3k}{(2\pi)^3}P_{\gamma}^i(k)\frac{d\Delta_{Q,
 iso}^2}{dz}  \,.
\end{eqnarray}
\begin{equation}
 \Delta_{Q,iso} (k) = {\sqrt{6} c_s \over \sqrt{2}} \left(k_{eq} \over k \right) e^{-k^2/k_D^2}.
\end{equation}

\begin{figure}
\begin{center}
\begin{tabular}{cc}
\includegraphics[width=0.45\textwidth]{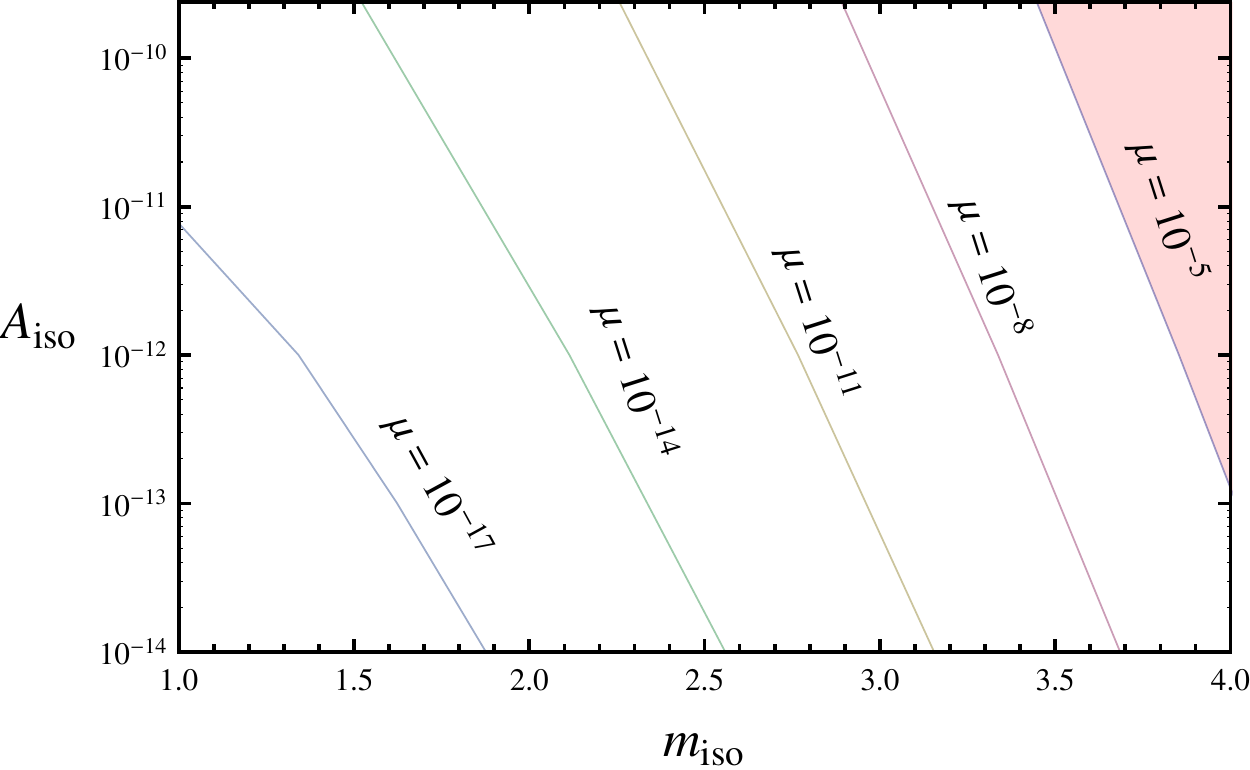}
\end{tabular}
\end{center}
\caption{Values of $\mu$ generated by energy injection due to the damping of
  acoustic oscillations as a function of the isocurvature
  spectral index, $m_{iso}$, for a range of values for the amplitude of the isocurvature
  mode, $A_{iso}$.  The maximum value shown for $A_{iso}$ corresponds to the upper bound determined by WMAP of $A_{iso}\lesssim 2.4\times 10^{-10}$.  The shaded region has been ruled out by the COBE-FIRAS instrument.}
\label{fig:muiso}
\end{figure}

\begin{figure}
\begin{center}
\begin{tabular}{cc}
\includegraphics[width=0.45\textwidth]{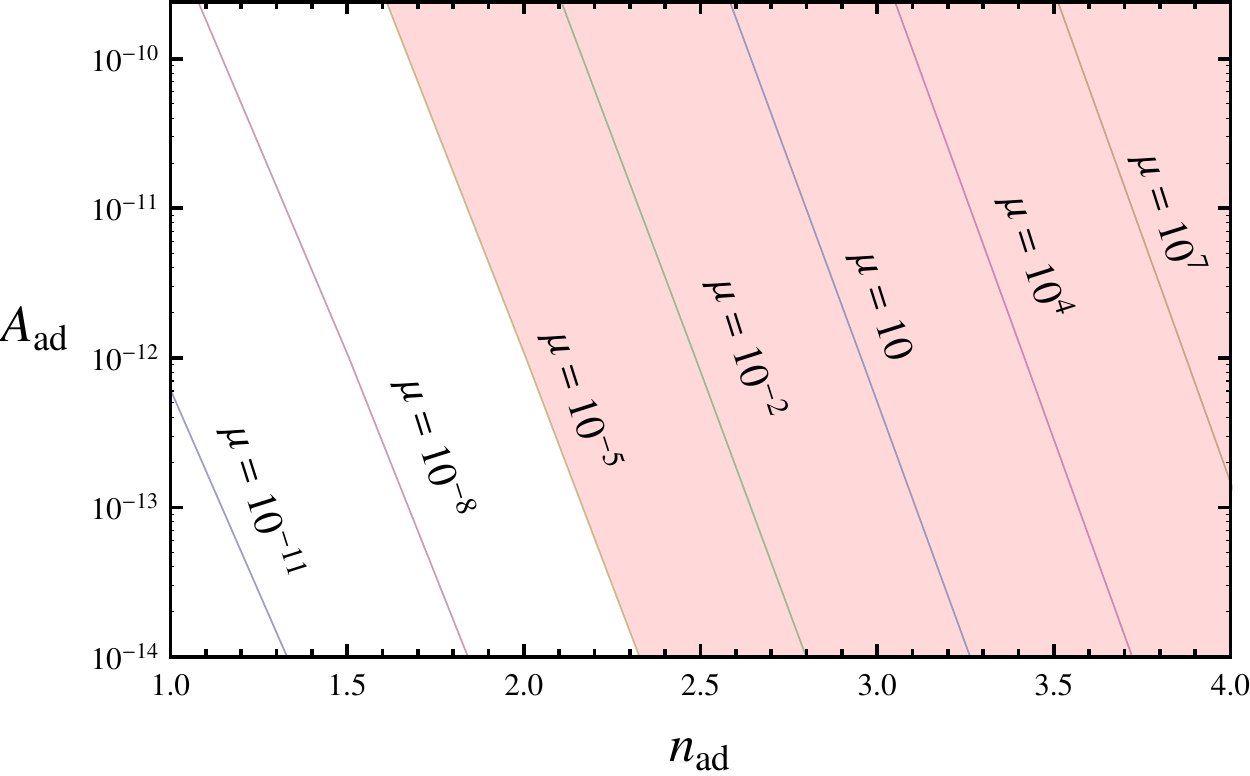}
\end{tabular}
\end{center}
\caption{Values of $\mu$ generated by energy injection due to the damping of
  acoustic oscillations is shown as a function of a generic adiabatic
  spectral index, $n_{ad}$, for a range of values for the amplitude of the adiabatic
  mode, $A_{ad}$.  This figure is shown for comparison with Fig.\ref{fig:muiso} in order to see the disparate contributions to $\mu$ from adiabatic and isocurvature modes.  The shaded region has been ruled out by the COBE-FIRAS instrument.}
\label{fig:muad}
\end{figure}

Integration of this expression, as in Eq.(\ref{mu}), will give the value
of $\mu$. In Fig.\ref{fig:muiso}, we show the results.
For comparison, we plot the analogous relevant quantities
for the adiabatic perturbations in Fig.\ref{fig:muad}.

Compared to the perturbations on large scales,
the perturbations  on small scales do not have enough time to  grow until they
re-enter the horizon. The density perturbations on small scales
are suppressed as shown Eq.~(\ref{eq:iso-delta}) and the energy of the
acoustic waves is  also small. As a result, $\mu$-distortion cannot
give as strong of a constraint on the isocurvature perturbations,  compared to
 the adiabatic perturbations.


%

With the introduction of possible isocurvature contributions to the
$\mu$-type distortion, one sees that a degeneracy between the various
sources of the distortion may arise.  This is due to the fact that $\mu$ is simply a scalar
quantity and there may be various contributions to its value.  Although there may be
correlations between isocurvature and adiabatic modes in various
inflationary models \cite{Langlois:1999dw} which could possibly
alleviate such a degeneracy, there may be other sources of isocurvature
modes, such as cosmic defects that can contribute to the degeneracy.
However, a future measurement of $\mu$ distortion, along with
complementary measurements of (or stronger bounds on) isocurvature
modes from upcoming experiments such as Planck, may break the
degeneracy or more strongly constrain the scenarios which can produce
the distortion.

\section{Discussion and Conclusions}

The dissipation of acoustic waves due to Silk damping creates
CMB spectral distortions of the blackbody spectrum. 
In this paper, we have calculated $\mu$-type distortions from energy injection into the CMB
due to the dissipation of acoustic 
waves from Silk damping, focusing on both the nearly-scale invariant 
adiabatic perturbations with a running spectral index, and isocurvature perturbations. 

\begin{figure}
\begin{center}
\begin{tabular}{cc}
\includegraphics[width=0.45\textwidth]{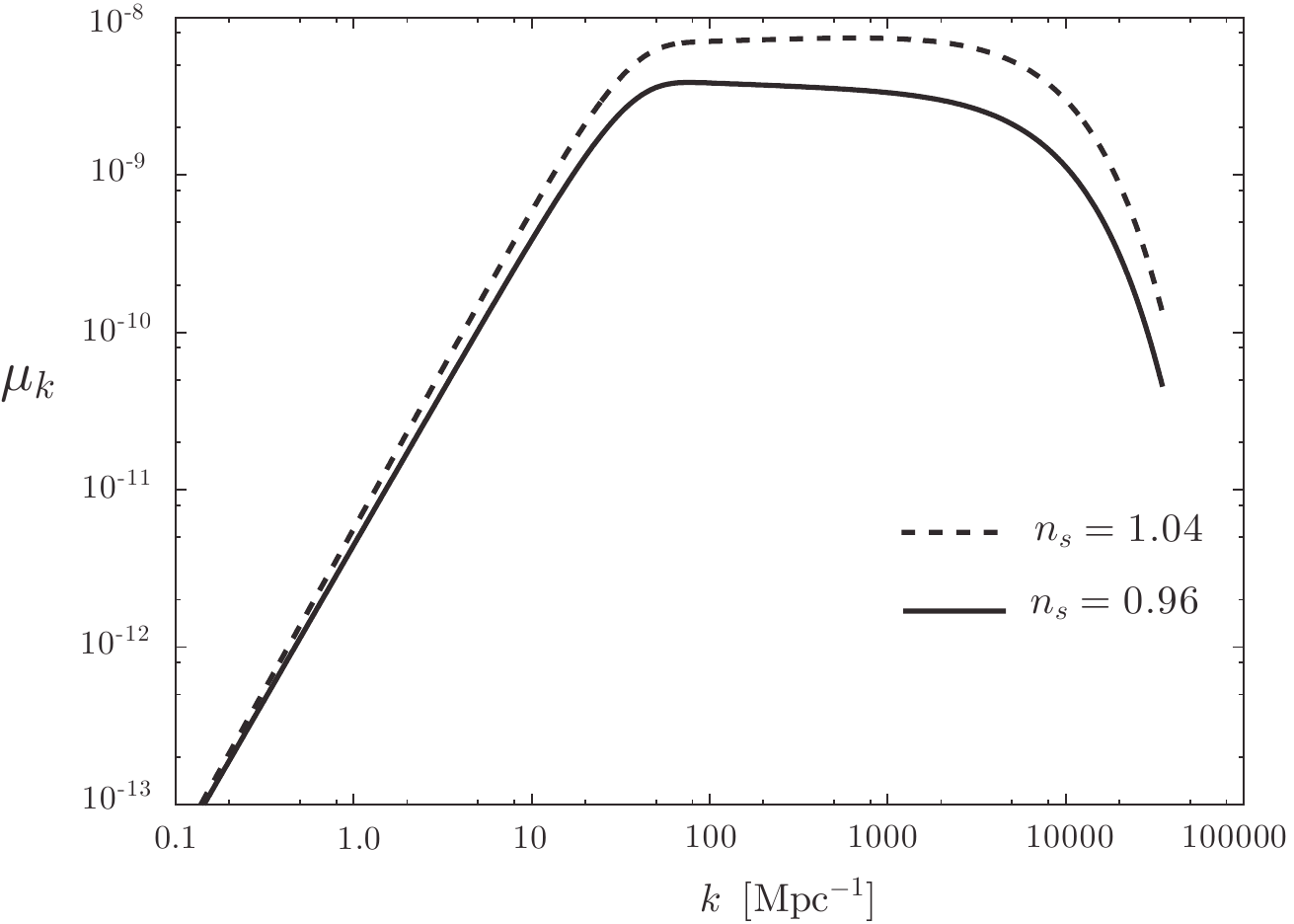}
\end{tabular}
\end{center}
\caption{The contribution to $\mu_k$ as a function of $k$. The chemical potential $\mu$ is given
by $\mu = 1.4 \int d \log k  \mu_k$.  One can see that the dominant contribution comes on scales of $k \sim 100~{\rm Mpc}^{-1}$, and decreases drastically on larger scales.}
\label{fig:kcontribution}
\end{figure}

Fig.~\ref{fig:mucontours} is our main result for $\mu$-distortion created
 by the adiabatic perturbations with a running spectral index.
Creation of this $\mu$-distortion depends on the integrated dissipation 
energy from the decoupling of the double Compton scatterings to
the decoupling of the thermalization by Compton scatterings.
We found that the dominant contribution comes from the 
dissipation energy at $k \sim 100~{\rm Mpc}^{-1}$ in the case of the 
nearly-scale invariant adiabatic perturbations, as one can see from Fig.~\ref{fig:kcontribution}. The power spectra,
which have the same amplitude at $k \sim 100~{\rm Mpc}^{-1}$,
generate the same $\mu$-distortion, even though each spectral index
and running index are different. As a result, a degeneracy arises between
$n_s$ and $\alpha_s$ as shown in Fig.~\ref{fig:mucontours}.
Although the current constraint on $\mu$-distortion by COBE FIRAS 
does not give a significant constraint on both $n_s$ and $\alpha_s$, 
the proposed PIXIE observer \cite{pixie} has the potential to give a much tighter constraint, 
regardless of detection or non-detection of $\mu$-distortion.
For example, the detection of $5 \times 10^{-5}$ $\mu$-distortion
by PIXIE will give the constraint, $-0.004 <\alpha_s<0$ for the 
scale invariant spectrum ($n_s=1$). This constraint is tighter 
than what is obtained by WMAP by nearly an order of magnitude. 
PLANCK will be able to give a
constraint comparable to PIXIE, although Planck's constraint
is the result of the measurement of the fluctuations on the scales
larger than $k\sim 1 {\rm Mpc}^{-1}$. Therefore, these measurements will be
complementary.

We also have evaluated $\mu$-distortion due to the isocurvature
perturbations. The isocurvature perturbations can grow outside the
horizon. Therefore, the smaller the scale of the perturbations, the
lower their amplitudes. As a result, $\mu$-distortions created by the
isocurvature perturbations are smaller compared to those generated by
the adiabatic perturbations with the same amplitude and spectral
index. We have found, considering the current constraint on the
isocurvature amplitude from WMAP, a nearly scale-invariant isocurvature
mode, which is motivated by the curvaton scenario, cannot produce
significant $\mu$-distortion.
Concerning possible cross-correlations between adiabatic and isocurvature
modes then, it is conjectured that,
even if a scale-invariant isocurvature mode is perfectly correlated
with the adiabatic mode, the correction to the $\mu$-distortion compared to that 
arising solely from the
adiabatic mode is very small.  The $\mu$-distortion measurements at the level 
proposed by PIXIE would then be inefficient in placing 
a constraint on the correlation between the scale-invariant adiabatic
and isocurvature modes.
However, PIXIE can place a strong constraint
on the amplitude of the isocurvature mode with a very blue spectrum ($n
< 2.5$), which can be generated, for example, in the axion model
\cite{kasuya:2009}.

Finally, we comment on other potential sources of $\mu$-distortion.
The $\mu$-distortion can be created by cooling of electrons,
which was first computed in \cite{Chluba:2011hw}, 
decaying particles \cite{hu-silk:1993}, dark matter annihilation \cite{McDonald:2000bk}, 
dissipation of the magnetic field 
energy \cite{jedamzik:2000}, Hawking radiation of
primordial black holes \cite{Tashiro:2008sf}, and cosmic strings \cite{Tashiro:2012nb}.
In particular, $\mu$-distortion due to the cooling by electrons is
predicted in the standard cosmology and enters with a negative sign.
Therefore, this distortion can cancel some or all of the positive $\mu$-distortion
due to the dissipation of acoustic waves. The predicted $\mu$-distortion is $-2 \times 10^{-9}$.

In the present work we have used approximate analytic forms for the transfer functions, and therefore a
more precise analysis is desirable.  Although we expect the size of the contributions from such
an analysis to be
negligible at the sensitivity of PIXIE, we require a more systematic 
analysis of $\mu$-distortion as PIXIE has the potential to give
a precise constraint on the primordial perturbations.
To address these issues, work is in progress to calculate the $\mu$-distortion more
precisely using numerical methods.

As was point out in \cite{Khatri:2011aj}, 
even a non-observation at the level of $\mu \sim \mathcal{O}(10^{-9})$ 
would be a sensitive cosmological probe.  Furthermore, it should also be noted that if a 
$\mu$-distortion \emph{were} observed on the order of $\sim 10^{-7}$ or greater, 
 the conventional cosmological picture would 
struggle to accommodate such a large value (see Fig.\ref{fig:mucontours}).
Thus, such an observation would lead to the tantalizing possibility of new physics\footnote{It has also recently been shown that $\mu$-distortions could probe new physics via tests of Gaussianity on very small scales \cite{Pajer:2012vz}.}.  Taken together these points underscore
the exciting insights that may emerge from future CMB blackbody measurements.  

\acknowledgements
It is a pleasure to thank Eiichiro Komatsu  and Tanmay Vachaspati for helpful discussions, and Thomas Jacques for comments on the manuscript.  
We would also like to thank Jens Chluba for his helpful comments and 
clarifications regarding the transfer functions used in the first version of this work. 
This research is supported in part by the Cosmology Initiative at Arizona State University and by the DOE.

\end{document}